\documentclass[aps,prd,twocolumn,nofootinbib,superscriptaddress]{revtex4}
\usepackage[hypertex]{hyperref}
\usepackage{graphicx}
\def\lesssim{\mathrel{\mathpalette\vereq<}}
\def\gtrsim{\mathrel{\mathpalette\vereq>}}

\newcommand{\be}{\begin{equation}}
\newcommand{\ee}{\end{equation}}
\newcommand{\bea}{\begin{eqnarray}}
\newcommand{\eea}{\end{eqnarray}}

\begin{document}

\pagestyle{plain}

\title{\begin{flushright} \texttt{UMD-PP-09-034} \end{flushright}
Probing Resonant Leptogenesis at the LHC}

\author{Steve Blanchet}
\author{Z. Chacko}
\author{Solomon S. Granor}
\author{Rabindra N. Mohapatra}
\affiliation{Department of Physics, University of Maryland, College
Park, MD, 20742}


\begin{abstract}

We explore direct collider probes of the resonant leptogenesis
mechanism for the origin of matter. We work in the context of
theories where the Standard Model is extended to include an
additional gauged U(1) symmetry broken at the TeV scale, and where
the light neutrinos obtain mass through a Type I seesaw at this
scale. The $C\!P$ asymmetry that generates the observed
matter-antimatter asymmetry manifests itself in a difference between
the number of positive and negative like-sign dileptons
$N(\ell^+\ell^+)-N(\ell^-\ell^-)$ that arise in the decay of the new
$Z'$ gauge boson to two right-handed neutrinos $N$, and their
subsequent decay to leptons. The relatively low efficiency of
resonant leptogenesis in this class of models implies that the
$C\!P$ asymmetry, $\varepsilon$, is required to be sizable, i.e. of
order one. In particular, from the sign of the baryon asymmetry of
the Universe, \emph{an excess of antileptons is predicted}. We
identify the domains in $M_{Z'}$--$M_N$ space where such a direct
test is possible and find that with 300~fb$^{-1}$ of data and no
excess found, the LHC can set the $2\sigma$ exclusion limit
$\varepsilon \lesssim 0.22$.

\end{abstract}

 \maketitle


\section{Introduction}

The origin of matter is a profound mystery. It is well known that the
primordial generation of a tiny baryon-antibaryon asymmetry can explain
why our present Universe consists almost exclusively of matter. The
possibility that this was put in ``by hand'' at the beginning is not
tenable since it is now generally believed that the universe underwent a
period of inflation, which would have diluted this initial amount to
negligible values. The observed asymmetry must therefore have been
generated after the end of inflation.

In 1967, Sakharov~\cite{sakharov} laid down the criteria under which a
baryon asymmetry can be spontaneously generated. Many particle physics
scenarios have subsequently been proposed that realize Sakharov's
conditions and thereby generate the observed matter-antimatter asymmetry.
In this paper, we focus on the mechanism of leptogenesis~\cite{fuku},
which is intimately tied to the origin of neutrino masses via the (Type I)
seesaw mechanism~\cite{seesaw}. The basic idea is that the heavy
right-handed (RH) Majorana neutrinos required for the seesaw mechanism can
produce an asymmetry between leptons and antileptons using the same
couplings that produce neutrino mass; this lepton asymmetry gets
transformed to a baryon asymmetry with the intervention of electroweak
sphaleron transitions, which are fast in the early
Universe~\cite{Kuzmin:1985mm}.

Unfortunately, in most generic versions of the leptogenesis scenario, the
RH neutrinos are superheavy and are therefore not accessible to colliders.
The situation, however, is very different if there is an additional U(1)
gauge symmetry broken at the TeV scale, under which the Standard Model
(SM) fields are charged. In general the gauge charges of the new U(1) will
forbid the $(LH)^2$ operator that generates Majorana neutrino mass. In
such a scenario the simplest possibility for neutrino mass generation
involves RH neutrinos at the TeV scale that carry charge under the
additional U(1), and which are necessary for anomaly cancellation. These
particles can only acquire Majorana masses once the U(1) symmetry is
broken, and are therefore required to be light.  The SM neutrinos have
small Yukawa couplings (of order the electron Yukawa coupling) to the
right-handed neutrinos, and acquire mass through a conventional Type I
seesaw. In this class of theories RH neutrinos can be pair produced
through decays of the $Z'$ associated with the new gauge symmetry. Their
subsequent decays $ N \rightarrow \ell^- W^+ $ and $N \rightarrow \ell^+
W^-$ constitute a window into the dynamics underlying neutrino mass
generation. In particular, the fact that the final state leptons can have
the same sign constitutes concrete evidence for the Majorana nature of
neutrinos. In this scenario, leptogenesis is possible provided that at
least two of the RH neutrinos are quasi-degenerate. This is the so-called
resonant leptogenesis mechanism~\cite{resonant}. Our considerations apply
to this case.

Let us define the $C\!P$ asymmetry parameter relevant for the LHC,
\begin{equation}\label{CPasym}
\varepsilon_{i}~=~\frac{\sum_{\alpha}\left[\Gamma(N_i\to \ell_{\alpha}^+ W^-)-\Gamma(N_i\to
\ell_{\alpha}^- W^+)\right]}{\sum_{\alpha} \left[\Gamma(N_i\to \ell_{\alpha}^+ W^-)+\Gamma(N_i\to
\ell_{\alpha}^- W^+)\right]},
\end{equation}
where $i=1,2,3$ and $\alpha=e,\mu,\tau$. The cosmological $C\!P$ asymmetry
is usually expressed in a somewhat different way, since the RH neutrino
can also decay into a $Z$ and a neutrino, or into a Higgs and a neutrino.
However, in the limit $M_N\gg M_{W^{\pm},Z,H}$ and for the self-energy
diagram which is the only one relevant for resonant leptogenesis, our
definition agrees with the conventional one.

As we discuss in Section III, leptogenesis at the weak scale is very
constrained in the class of models we consider, because of the
$Z'$-mediated scattering processes, $f\bar{f}\leftrightarrow NN$. The Type
III seesaw case exhibits similar behavior~\cite{Hambye}. In fact, it is
non-trivial that an allowed region in the space ($M_{Z'}$, $M_N$) exists
at all~\cite{Frere:2008ct}. The final baryon asymmetry is given by
\begin{equation}\label{etaB0}
\eta_B\simeq 10^{-2}\varepsilon \,\kappa^{\rm fin}
\end{equation}
In our scenario, the efficiency factor at the end of leptogenesis,
$\kappa^{\rm fin}$, is of order $10^{-7}$--$10^{-8}$ for $Z'$ masses
accessible at the LHC. It then follows that the $C\!P$ asymmetry parameter
$\varepsilon$ must be of order one in order to match the observed baryon
abundance, $\eta_B=(6.2\pm 0.15)\times 10^{-10}$~\cite{Komatsu:2008hk}.
This can be achieved if the RH neutrinos are degenerate to one part in
$10^{14}$~\cite{resonant}. An example of a simple framework in which such
a spectrum of neutrinos can naturally arise is shown in Appendix B.
Consequently, if the RH neutrinos satisfy the kinematic requirement that
$M_{Z'}\geq 2 M_N$, so that the decay $Z'\to NN$ is allowed, then $N$ must
decay into leptons with order one asymmetry if leptogenesis is indeed at
the origin of the observed baryon asymmetry. This ``large'' value of
$\varepsilon$ then allows the number of positive like-sign dilepton,
$N(\ell_{\alpha}^+\ell_{\beta}^+)$, to be significantly different from the
negative like-sign ones, $N(\ell_{\alpha}^-\ell_{\beta}^-)$. This
difference directly measures $(2\sum_i \varepsilon_i) /(\sum_i 1)$, as we
discuss in Section IV. Therefore, an observation of this quantity
constitutes a direct test of TeV-scale leptogenesis in this class of
models. Specifically, as expected from leptogenesis, an excess of
antileptons over leptons at the LHC is predicted by the sign of the baryon
asymmetry of the universe. It should be emphasized here that the fact that
large $C\!P$ asymmetries are required is entirely due to the presence of
the new $Z'$. In the standard resonant scenario at TeV scale, $C\!P$
asymmetries of order $10^{-4}$ suffice, which are much too small to be
observed at colliders.

\section{Determining the Baryon Asymmetry}

We consider the addition of an additional Abelian gauge group to the SM. For concreteness, we
will take this new U(1) to be $B - L$~\cite{marshak}. An alternative
choice will not significantly affect our
conclusions. The Lagrangian of this model differs from the SM by the usual Type I seesaw term,
\begin{equation}
\mathcal{L}\supset {\rm i}\overline{N_{Ri}}D_{\mu}\gamma^{\mu} N_{Ri} -h_{\alpha i}
\overline{L_{L\alpha}}\tilde{\Phi}
N_{R i } -{1\over 2} M_{N i} N_{Ri}^T C N_{Ri} +h.c.,
\end{equation}
with $i=1,2,3$, $\alpha=e,\mu,\tau$, $L$ and $\Phi$ are $SU(2)$ doublets, $\tilde{\Phi}={\rm
i} \sigma_2 \Phi^*$, and $D_{\mu}=\partial_{\mu} -{\rm i} g_1'Y_{B-L} B'_{\mu}$.
The charges under this group are particularly simple: $Y(Q_L)=Y(D_R)=Y(U_R)=1/3$ and
$Y(L_L)=Y(E_R)=Y(N_R)=-1$, for quarks and leptons, respectively.

The efficiency factor $\kappa^{\rm fin}$ introduced above is determined by solving numerically
the set of Boltzmann equations relevant for this model (see for
instance~\cite{Racker:2008hp}). In comparison to the standard Type I case, there is an
additional scattering term in the equation for the evolution of the $N_i$ number density. In
order to have enough $C\!P$ asymmetry when $M_N\sim 1$~TeV, the RH neutrinos need to be
degenerate to a high degree~\cite{resonant}. However, in the computation of the efficiency
factor the small mass differences do not matter, and we can assume $M_{N_i}\equiv M_N$,
$i=1,2,3$. Moreover, since leptogenesis occurs in the TeV range, flavor
effects~\cite{Barbieri:1999ma} must be included, and the three flavors are
distinguished. Including flavor effects and the contributions from all RH neutrinos, we can
express the final baryon asymmetry produced through leptogenesis as~\cite{buch}
\begin{equation}\label{etaB}
\eta_B\simeq 10^{-2}\sum_{\alpha}N_{\Delta_{\alpha}}(z\to \infty)\simeq 10^{-2}\sum_{i,\alpha} \varepsilon_{i\alpha} \,
\kappa_{i\alpha}(z\to\infty),
\end{equation}
where $z=M_{N}/T$, $\Delta_{\alpha}=B/3-L_{\alpha}$ and $\sum_{\alpha}\varepsilon_{i\alpha}=\varepsilon_i$. The
conveniently normalized number density $N_{\Delta_{\alpha}}$ or the efficiency
factor $\kappa_{i\alpha}$ are found solving the relevant set of Boltzmann equations.
Including only the dominant processes, i.e. decays, inverse decays, as well as scatterings mediated by $Z'$,
the latter are given by
\begin{eqnarray}
{{\rm d}N_{N_i}\over {\rm d}z}=-D(K_i)(N_{N_i}-N_{N_i}^{\rm eq})-2\,S_{Z'}\left(N_{N_i}^2-(N_{N_i}^{\rm eq})^2\right)\; , \nonumber\\
{{\rm d}N_{\Delta_{\alpha}}\over {\rm d}z}=\sum_i \varepsilon_{i\alpha} \, D(K_i)\, (N_{N_i}-N_{N_i}^{\rm eq})-\sum_i
W^{\rm ID}(K_{i\alpha}) N_{\Delta_{\alpha}}\; ,\nonumber
\end{eqnarray}
where $N_{N_i}^{\rm eq}(z)={1\over 2} z^2 \mathcal{K}_2(z)$, $D(K,z)=K z
\mathcal{K}_1(z)/\mathcal{K}_2(z)$ and $W^{\rm ID}(K,z)={1\over 4}K\mathcal{K}_1(z)z^3$, with
$\mathcal{K}_i(z)$ being the modified Bessel function of the $i$th type.
Since RH neutrinos in our model track closely equilibrium, it is possible to use the approximation
${\rm d}N_{N_i}/ {\rm d}z \simeq {\rm d}N_{N_i}^{\rm eq}/ {\rm d}z$~\cite{buch} to write the efficiency factor
$\kappa_{i\alpha}$ as
\begin{eqnarray}\label{efficiency}
\kappa_{i\alpha} (z,z_{\rm in})&\simeq& \int^z_{z_{\rm in}}{\rm d}z'
{{\rm d}N_{N_i}^{\rm eq}\over {\rm d}z'}
\frac{D(K_i,z')}{D(K_i,z')+4S_{Z'}(z')N_{N_i}^{\rm eq}(z')}\nonumber\\
&&\times\exp \left(-\int^{z}_{z'} \sum_i W^{\rm ID}(K_{i\alpha},z'') {\rm
d}z''\right),
\end{eqnarray}
The flavored decay
parameter is given by the ratio of the decay width to the Hubble expansion when the mass equals
the temperature,
\begin{equation}\label{decpar}
K_{i\alpha}={\widetilde{\Gamma}_{\rm D} (N_i\to L_{\alpha} \Phi +\bar{L}_{\alpha}
\Phi^{\dagger})\over H(z=1)}={|h_{\alpha i}|^2 v^2\over M_N m_{\star}},
\end{equation}
with $m_{\star}=1.08\times 10^{-3}$~eV. Summing over alpha gives the total decay parameter
$K_i=(h^{\dagger}h)_{ii}v^2/(M_N m_{\star})$. It is useful to define two typical values of the
decay parameter deduced from neutrino masses:
$K_{\rm sol}\equiv m_{\rm sol}/m_{\star}\simeq 8.1$ and $K_{\rm atm}\equiv m_{\rm
atm}/m_{\star}\simeq 46$.
Note how in general the efficiency factor in Eq.~(\ref{efficiency}) depends on both
$K_i$ and $\sum_i K_{i\alpha}$,
and not only the sum $\sum_i K_{i\alpha}$ as in the usual resonant Type I
case~\cite{Blanchet:2006dq}.
The scattering rate $S_{Z'}\equiv
\gamma_{Z'}/(H n_N^{\rm eq} z)$, where $n_N^{\rm eq}$ is the RH neutrino equilibrium
number density, and $\gamma_{Z'}$ is a reaction density which depends
on the following reduced cross section\footnote{Our result agrees with~\cite{Plumacher:1996kc} but
disagrees with~\cite{Racker:2008hp}.}:
\begin{equation}
\hat{\sigma}_{Z'}(x)={13 g_1'^4\over 6 \pi} {\sqrt{x(x-4)^3}\over (x-M_{Z'}^2/M_N^2)^2+
M_{Z'}^2\Gamma_{Z'}^2/M_N^4},
\end{equation}
where $x=s/M_N^2$. The total $Z'$ decay width in this model is given by
\begin{equation}
\Gamma_{Z'}={g_1'^2\over 24 \pi}M_{Z'}\left(13+
3(1-4M_N^2/M_{Z'}^2)^{3/2}\right).
\end{equation}
If one were to plot $S_{Z'}(z)$ and $D(z)$, one would immediately see that $S_{Z'}\gg D$ for
$z\ll 1$, implying that essentially no asymmetry is produced at high temperatures $T\gg M_N$.
The asymmetry is created once the Boltzmann suppression in $N_{N_i}^{\rm eq}$ starts acting,
when $T\lesssim M_N$. It turns out that the maximal efficiency occurs at very large values of
$K$, of the order of $10^3$--$10^4$~\cite{Frere:2008ct}. We will be more conservative, and
simply assume values of $\sum_i K_{i\alpha}$ that are motivated by neutrino masses, and for
definiteness further assume that $K_{i \alpha} = K_i/3$ for each flavor $\alpha$, except in
the case of normal hierarchy, where the washout in the $e$ flavor is typically
suppressed~\cite{Blanchet:2008zg}. Note that both the assumption of flavor universality and
$K\sim m_{\nu}/m_{\star}$ are conservative in the sense that relaxing them, we would get
(slightly) larger efficiency factors. Since we know that $\sum_i K_i> K_{\rm sol}+K_{\rm
atm}\,(2\,K_{\rm atm})$ for normal (inverted) hierarchy, and $\sum_i K_i>300$ if $m_1\simeq
m_2\simeq m_3\simeq 0.1$~eV, i.e. for a quasi-degenerate spectrum, we will consider the
following three benchmark points: $\sum_i K_{i\tau,\mu}= 25,\sum_i K_{ie}=5$ for normal
hierarchy, $\sum_i K_{i\alpha}= 30$ for inverted hierarchy, and finally $\sum_i K_{i\alpha}=
100$ for a quasi-degenerate spectrum. With reasonable assumptions about the flavored $C\!P$
asymmetries $\varepsilon_{i\alpha}$, it turns out that the normal hierarchy and inverted
hierarchy cases lead to very similar results. This is because of the weak dependence of the
final efficiency factor on $\sum_{i}K_{i\alpha}$. In what follows we therefore present the
results for these two cases together.

\begin{figure}[t]
\includegraphics[width=0.3\textwidth]{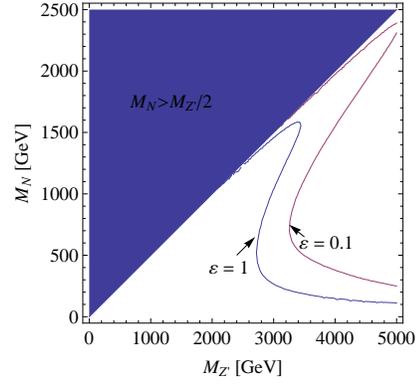}
\caption{Regions in the space $(M_{Z'}$--$M_{N})$ where leptogenesis
can be tested for the case of normal or inverted hierarchy. The regions to the
right and above the colored curves are allowed.}
\label{fig:1}
\end{figure}

\begin{figure}[t]
\includegraphics[width=0.3\textwidth]{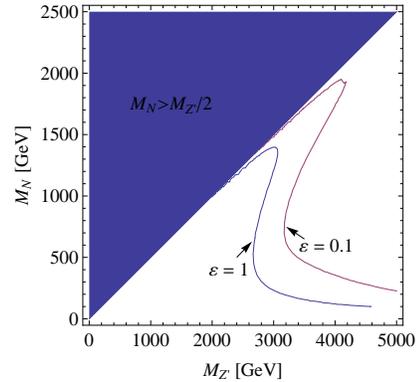}
\caption{Same as Fig.~\ref{fig:1} but for the case of quasi-degenerate neutrinos.}
\label{fig:2}
\end{figure}

We have numerically integrated Eq.~(\ref{efficiency}), and assumed for concreteness that
$\varepsilon_1=\varepsilon_2=\varepsilon_3 \equiv \varepsilon$ and $K_1=K_2=K_3$,
in order to
get a typical region in the plane $M_{Z'}$--$M_N$ where leptogenesis is
successful. We have assumed that the production of
asymmetry stops immediately once $T<T_{\rm sph}$, the sphaleron
freeze-out temperature. For a Higgs mass of 120~GeV, this is given
by 130~GeV~\cite{Burnier}. The results are shown in Figs.~\ref{fig:1}
and \ref{fig:2} for the value of the new
gauge coupling $g'_1=0.2$. The allowed regions are to the right
and above the colored lines. Inside the contour of $\varepsilon = 1$,
the efficiency factor is $\kappa^{\rm fin}(\infty)\gtrsim 10^{-8}$, and
inside $\varepsilon = 0.1$, the efficiency factor calculated is
$\kappa^{\rm fin}(\infty)\gtrsim 10^{-7}$. As mentioned above, we are showing only
one plot for the normal and inverted
hierarchy cases because the allowed regions are almost identical. We have restricted the plane
to
$M_{Z'}\leq 5$~TeV and $M_N\leq M_{Z'}/2$, which is favored for discovery at the LHC.
Note however that leptogenesis is also successful
in the region $M_N\geq M_{Z'}/2$, as shown in~\cite{Frere:2008ct}.
As pointed out earlier, the efficiency factor is maximal at large values
of $K$. This upper bound implies an absolute lower bound for the $Z'$ mass
in order to have successful leptogenesis: $M_{Z'}>2.6~(2.1)$~TeV for
$g_1'=0.2~(0.1)$. For smaller values, a $C\!P$ asymmetry parameter greater
than one would be required, which is unphysical. Therefore, if a $Z'$ with
a mass below 2~TeV is discovered at the LHC, and RH neutrinos are observed
with masses below $M_{Z'}/2$, then leptogenesis is not possible, and some
alternative mechanism of baryogenesis must be present. In any such
scenario, the bounds on any pre-existing asymmetry derived
in~\cite{Blanchet:2008zg} must be taken into account.

\section{Experimental prospects}

\begin{figure}[t]
\includegraphics[width=0.4\textwidth]{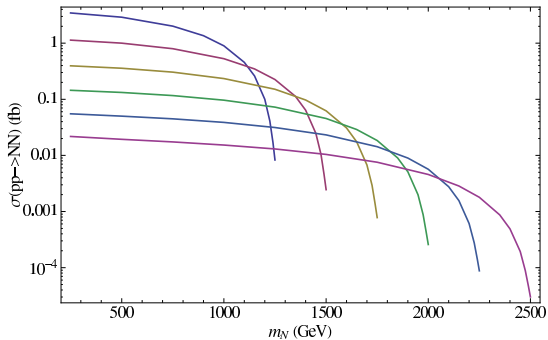} \caption{Total
cross section $pp\to Z'\to NN$ for $g_1'=0.2$ and varying $M_{Z'}$ between
2.5 and 5~TeV in steps of 500~GeV (top to bottom).} \label{fig:3}
 \end{figure}

We show in Fig.~\ref{fig:3} the total LHC cross section calculated
using CalcHEP~\cite{calchep} at 14 TeV to any pair of RH neutrinos,
$pp\to Z'\to NN$~\cite{rizzo}. We have fixed $g_1'=0.2$ and varied
$M_{Z'}$ between 2.5 and 5~TeV in steps of 500~GeV. For
$M_{Z'}=3$~TeV and $M_N~=~500$~GeV, we see that we obtain a total
cross-section of about 1~fb, corresponding to about 300 signal
events with 300 fb$^{-1}$ of data. With 1000 fb$^{-1}$ of data this
increases to 1000 signal events.
The decay modes of the RH neutrino that are relevant for us are $N\to
\ell^{\pm}W^{\mp}$, which constitute half of the total decay rate of each
RH neutrino in the limit $M_N\gg M_{W^{\pm},Z,H}$, as a consequence of the
Goldstone boson equivalence theorem. We will concentrate on
events where both right-handed neutrinos decay to charged leptons, since
the backgrounds associated with such events tend to be smaller.  To first
order in $\varepsilon$, the asymmetry between positive and negative
like-sign dileptons is given by
 \begin{equation}
\label{observable}
{N(\ell^+\ell^+)-N(\ell^-\ell^-)\over N(\ell^+\ell^+)+N(\ell^-\ell^-)}=
{2\sum_i\varepsilon_i \over \sum_i 1},
 \end{equation}
where we sum over all RH neutrino contributions. Therefore the difference between
positive and negative like-sign dilepton events provides a direct probe of the asymmetry
parameter.

We will primarily focus on events where both $W$ bosons decay
hadronically. This has the advantage of avoiding ambiguities that
can arise in distinguishing the leptons arising directly from RH
neutrino decay from those arising subsequently from $W$ boson decay.
The formula Eqn.~({\ref{observable}}) that relates the asymmetry
parameter to the number of dilepton events is unaffected by this
restriction. With 300 fb$^{-1}$ and no asymmetry, the expected
number of such like-sign dileptons is
$N(\ell^+\ell^+)=N(\ell^-\ell^-)= 8.3 \pm 2.0$ at 1$\sigma$ for
$M_{Z'}=3$~TeV
 and $M_N=500$~GeV. With
1000 fb$^{-1}$ this goes up to $27.8 \pm 3.7$. If we assume
$\varepsilon_1= \varepsilon_2=\varepsilon_3 \equiv \varepsilon$, as
we did for the leptogenesis analysis, and further that all such
events can be identified and distinguished from background, we
estimate that with 300 fb$^{-1}$ of data and no excess observed, the
LHC will be able to set a 2$\sigma$ exclusion limit of order
$\varepsilon<0.22$. With 1000 fb$^{-1}$ of data, this improves to
$\varepsilon<0.13$. However, these assumptions clearly correspond to
a best-case scenario, and a careful analysis that incorporates the
effects of backgrounds, the acceptance of the detector and the
challenge of signal identification in the LHC environment is
required before a firm conclusion can be drawn. In this paper, we
will limit ourselves to estimating the leading backgrounds, leaving
a more complete study of these effects for future work.

Since the invariant mass in each event is so large, the dominant background at the LHC
arises from events involving a top and anti-top that each decay leptonically, and
where the charge of one of the resulting leptons is misidentified. To estimate this we
calculate using CalcHEP the production cross section at the LHC for a $t \bar{t}$ pair
with a total invariant mass of 3 TeV or more, and find that it is of order 10 fb.
Requiring that this pair decay leptonically only reduces this to about a fb, and so
additional cuts are needed. The cuts to be used depend on whether any of the leptons
in the event is a tau. In events where all the leptons are electrons or muons, the
requirement that the charge of one of the leptons is misidentified reduces this
background to about 0.02 fb, well below the level of the signal. In addition, since
these signal events involve no missing energy, this can be used to reduce the
background. Finally, requiring that the invariant mass of each $\ell W$ pair add up to
the mass of the $N$ provides another very strong constraint. We conclude that this
background is under control.

Background events involving one or more taus are more difficult to constrain, since
tau charge identification is in general less reliable unless the tau decays
leptonically. In addition, since tau events always involve missing energy, such a cut
cannot be used. In events where at least one of the leptons is an electron or a muon
(about $90\%$ of events) requiring that the invariant mass of this $\ell W$ pair add
up to the mass of the $N$ can be used to reduce the background. Furthermore, the fact
that the direction of the invisible decay products of a highly boosted tau will align
almost exactly with that of the visible decay products leads to a constraint on the
direction of the missing energy in the event, and also to a separate invariant mass
constraint. The same fact can be used to restrict the background even in events where
both the leptons are taus. Separately, the fact that the mass of the RH neutrino is in
general much larger than the mass of the top quark means that the angular
distributions of their decay products are very different, leading to additional
constraints.

Another possible background at the LHC arises from $j j W^{+} W^{+}$, with the $W$'s
decaying leptonically. We have studied this background using
MadGraph/MadEvent~\cite{madgraph}. Requiring that the energy of the visible particles
in the event be greater than 3 TeV, that the transverse momentum of each lepton be at
least 100 GeV, and that the total missing energy in the event be less than 50 GeV
reduces this background to below 0.005 fb. We have verified that these cuts do not
significantly affect the signal. An identical set of cuts can be used to kill the
background arising from $j j W^+ W^-$, with the charge of one of the leptons
misidentified. This analysis suggests that the backgrounds are under control, though
further study is required before a firm conclusion can be drawn.

While we have focused on events where both $W$ bosons decay into
hadrons, events where one $W$ boson decays leptonically can also be
used to extract information about the $C\!P$ asymmetry parameter
$\varepsilon$, provided all the leptons in the event are electrons
or muons. All the missing energy in such events is associated with a
single neutrino, allowing the corresponding $W$ boson to be
reconstructed. The lepton arising from decay of the $W$ can then be
identified and distinguished from the leptons arising directly from
RH neutrino decay. The primary background to such events arises from
$j j W Z$, and can be made negligible after cuts.

It is also possible to calculate $\varepsilon$ from events where only one of the RH
neutrinos decays to a charged lepton and $W$, while the other decays to a neutrino and
$Z$, or alternatively to a neutrino and Higgs. Although the SM backgrounds are
potentially larger, there are more such events than like-sign dilepton events.
$\varepsilon$ is related to the asymmetry in the number of events with positively
charged leptons relative to negatively charged leptons.
 \begin{equation}
\label{observable2}
{N(\ell^+)-N(\ell^-)\over N(\ell^+)+N(\ell^-)}=
{\sum_i\varepsilon_i \over \sum_i 1},
 \end{equation}
Therefore, this class of events can be used as an independent measure of
the value of $\varepsilon$.

In order to achieve order one values for the $C\!P$ asymmetry, we must be
close to the resonance region $M_j-M_i\sim \Gamma/2$, where interference
effects may be important. However, for perturbation theory in the
computation of the $C\!P$ asymmetry to be applicable, we require
$M_j-M_i\gtrsim 3 \, \Gamma/2$ ~\cite{Anisimov:2005hr}, which corresponds
to $\varepsilon\lesssim 0.3$.  In Appendix A, we explicitly verify that
interference effects arise only at order $\varepsilon^3$ for the $C\!P$
violating observable in Eq.~(\ref{observable}), and at order
$\varepsilon^2$ in the rates. Then, for $\varepsilon\lesssim 0.3$, the
corrections to our results from interference are not more than 10\%.

An important element for resonant leptogenesis is the presence of at least
two degenerate RH neutrinos. The extreme degeneracy in their masses
implies that it will not be possible to determine the number of RH
neutrinos based on invariant mass measurements. Nevertheless, by measuring
their branching ratios into leptons of various flavors, it may be possible
to distinguish the cases of one, two and three RH neutrinos, even in the
absence of any observed $C\!P$ asymmetry. The decay probability of one RH
neutrino into a certain lepton flavor (either lepton or antilepton) is
given by
\begin{equation}
P_{i\alpha}={|h_{\alpha i}|^2 \over (h^{\dagger}h)_{ii}}.
\end{equation}
Clearly the sum of the probabilities must equal one: $\sum_{\alpha} P_{i\alpha}=1$, for
$i=1,2$ and 3. Then, the probability of a given dilepton event to involve the flavors
$\alpha$ and $\beta$, which
can be directly measured at
the LHC, is
\begin{equation}
\label{palphabeta}
P(\ell_{\alpha}\ell_{\beta})={\sum_i P_{i\alpha}P_{i\beta}\over \sum_i 1},
\end{equation}
where $\alpha,\beta=e,\mu,\tau$, and $i$ runs over the RH neutrinos (1, 2 or 3). Note that
the tree level expressions used here are enough for our purposes since the total rates into leptons
plus antileptons are only corrected by $\epsilon^2$ as described above, contrary to the difference
in rates into leptons versus antileptons, which goes like $\varepsilon$. We have the
additional constraint that $\sum_{\alpha,\beta}P(\ell_{\alpha}\ell_{\beta})=1$, which implies
that one of the six equations in Eq.~(\ref{palphabeta}) is redundant. With only one RH
neutrino, say $N_1$, we have five equations for two unknowns, $P_{1e}$ and $P_{1\mu}$
($P_{1\tau}$ is known from the sum of probabilities), which means that the system is highly
overconstrained. If no consistent solution to these five equations can be found, it means that
there must be more than one RH neutrino. If there are two RH neutrinos, say $N_1$ and $N_2$,
we have five equations for the four unknowns $P_{1e}$, $P_{1\mu}$, $P_{2e}$ and $P_{2\mu}$ and
so the system is still overconstrained. Therefore this case can potentially also be
distinguished from that of three RH neutrinos.

\section{Conclusion}

We have shown that in a model with TeV-scale RH neutrinos and $Z'$ gauge
boson, resonant leptogenesis is possible, and requires a large (order one)
$C\!P$ asymmetry to work. The allowed range for leptogenesis in the space
$M_{Z'}$--$M_N$ is very constrained in the LHC-favored situation
$M_N<M_{Z'}/2$, and favors larger values of the $Z'$ mass, $M_{Z'}>
2$~TeV. The large $C\!P$ asymmetry required in the decay of the RH
neutrinos may have observable consequences at the LHC, in particular an
asymmetry in the number of positive and negative like-sign dilepton
events. Specifically, the sign of the baryon asymmetry of the Universe
implies an excess of anti-leptons over leptons. If no excess is observed,
we find that with 300~fb$^{-1}$ of integrated luminosity the LHC will be
able to exclude at 2$\sigma$ that $\varepsilon \gtrsim 0.22$. Finally,
although the RH neutrino masses are essentially identical, their couplings
to leptons are not, and we show that some simple linear algebra
considerations allow us to distinguish the cases of one, two and three
degenerate RH neutrinos even in the absence of any observed $C\!P$
asymmetry.

\acknowledgments
It is a pleasure to thank S. Eno and C. Kilic for useful comments.
ZC is supported by the NSF under grant PHY-0801323. RNM is supported by the NSF under grant
PHY-0652363.

\appendix
\section{Interference effects}

In the following we want to compute the parameter dependence of the interference terms in
the process $Z'\to \ell_{\alpha} \Phi \ell_{\beta}\Phi$, where RH neutrinos are exchanged
in the intermediate step.

\subsection{Field theory derivation}
The amplitude for the process
$Z'(p_{Z'})\to \ell_{\alpha}(k_1) \Phi(k_2) \ell_{\beta}(k_1')
\Phi(k_2')$ is given by
\begin{eqnarray}
\mathcal{M}(\ell_{\alpha}\ell_{\beta})&=&g_1'\bar{u}_{\alpha}(k_1)P_R
[h_{\alpha 1} S_{11}(p)+h_{\alpha 2} S_{12}(p)]\gamma_{\mu}
\gamma_5 \nonumber \\
&&\varepsilon^{\mu}(p_{Z'})[h_{\beta 1}S_{11}(p')+h_{\beta 2}S_{12}(p')] P_R v_{\beta}(k_1')\nonumber\\
&+&g_1'\bar{u}_{\alpha}(k_1)P_R[h_{\alpha 1} S_{21}(p)+h_{\alpha 2}
S_{22}(p)]\gamma_{\mu} \gamma_5 \nonumber \\
&&\hspace{-0.5cm}\varepsilon^{\mu}(p_{Z'})[h_{\beta
1}S_{21}(p')+h_{\beta 2}S_{22}(p')] P_R
v_{\beta}(k_1').\label{amplitude}
\end{eqnarray}
Following \cite{Anisimov:2005hr} we decompose the propagator into chiral
components
\begin{eqnarray}
S_{ij}(p)&=&P_R S_{ij}^{RR}(p^2)+P_L S_{ij}^{LL}(p^2) \nonumber\\
&+&P_L \displaystyle{\not}p S_{ij}^{LR}(p^2)+P_R \displaystyle{\not}p S_{ij}^{RL}(p^2).
\end{eqnarray}
It can then be easily shown that the only non-zero components are given by
\begin{equation}
S^{RR}(p^2)[\ldots]p\!\!/' S^{LR}(p'^2) + \displaystyle{\not}p S^{RL}(p^2) [\ldots] S^{RR}(p'^2).
\end{equation}
The renormalized and resummed matrices of propagators $S^{RR}$, $S^{LL}$, $S^{LR}$ and $S^{RL}$ can be found
in~\cite{Anisimov:2005hr} and will not be reproduced here. The crucial parameters for the following
discussion are the poles of the propagators, given to leading order in the small Yukawa couplings
by~\cite{Anisimov:2005hr}
\begin{equation}\label{poles}
s_i\simeq M_i^2-iM_i\Gamma_i\,.
\end{equation}

We have to evaluate now the modulus squared of the amplitude (\ref{amplitude}). We will omit for
clarity the spinor and gamma matrices product since they will be common to all
terms\footnote{It is given by $|\bar{u}_{\alpha} P_R \displaystyle{\not}p \gamma_{\mu}\gamma_5 \varepsilon^{\mu} P_R v_{\beta}
+\bar{u}_{\alpha} P_R  \gamma_{\mu}\gamma_5 \varepsilon^{\mu}p\!\!/' P_R v_{\beta}|^2$.}. We obtain
\begin{eqnarray}
|\mathcal{M}|^2/g_1'^2&=& \nonumber\\
&&\hspace{-1.5cm}|h_{\alpha 1} h_{\beta 1} S_{11}(p) S_{11}(p')+h_{\alpha 1} h_{\beta 2} S_{11}(p) S_{12}(p')\nonumber \\
&&\hspace{-1.5cm}+h_{\alpha 2} h_{\beta 1} S_{12}(p) S_{11}(p')+h_{\alpha 2} h_{\beta 2} S_{12}(p) S_{12}(p')|^2\nonumber \\
&&\hspace{-1.5cm}+ |h_{\alpha 1} h_{\beta 1} S_{21}(p) S_{21}(p')+h_{\alpha 1} h_{\beta 2} S_{21}(p) S_{22}(p')\nonumber\\
&&\hspace{-1.5cm}+h_{\alpha 2} h_{\beta 1} S_{22}(p) S_{21}(p')+h_{\alpha 2} h_{\beta 2} S_{22}(p) S_{22}(p')|^2\nonumber \\
&&\hspace{-2cm}+2{\rm Re}\left[\left(h_{\alpha 1} h_{\beta 1} S_{11}(p) S_{11}(p')+h_{\alpha 1} h_{\beta 2} S_{11}(p) S_{12}(p')\right.\right.\nonumber\\
&&\hspace{-1.5cm}\left.\left.+h_{\alpha 2} h_{\beta 1} S_{12}(p) S_{11}(p')+ h_{\alpha 2} h_{\beta 2} S_{12}(p) S_{12}(p') \right)^{\star}\right.\nonumber\\
&&\hspace{-1.5cm}\left.\times\left(h_{\alpha 1} h_{\beta 1} S_{21}(p) S_{21}(p')+h_{\alpha 1} h_{\beta 2} S_{21}(p) S_{22}(p')\right.\right.\nonumber \\
&&\hspace{-1.5cm}\left.\left.+h_{\alpha 2} h_{\beta 1} S_{22}(p) S_{21}(p')+h_{\alpha 2} h_{\beta 2} S_{22}(p)
S_{22}(p')\right)\right].
\end{eqnarray}
Note that we omit the upper scripts $RR$, $RL$ or $LR$ because the distinction will become irrelevant
on mass shell. Keeping  only first order terms in the off-diagonal elements of the propagator matrix
$S_{12}=S_{21}$ and we obtain
\begin{eqnarray}
|\mathcal{M}|^2/g_1'^2&=& |h_{\alpha 1}|^2 |h_{\beta 1}|^2 |S_{11}(p)|^2 |S_{11}(p')|^2\nonumber\\
&+&|h_{\alpha 2}|^2|h_{\beta 2}|^2 |S_{22}(p)|^2 |S_{22}(p')|^2\nonumber\\
&+& 2{\rm Re}\left(h_{\alpha 1}^{\star} h_{\beta 1}^{\star} h_{\alpha 2} h_{\beta 2} S_{11}^{\star}(p)S_{22}(p) S_{11}^{\star}(p')
S^{\star}_{22}(p')\right)\nonumber\\
&+& 2 |h_{\alpha 1}|^2 |S_{11}(p)|^2{\rm Re}\left(h_{\beta 1}^{\star} h_{\beta 2} S_{11}^{\star}(p') S_{12}(p')\right)\nonumber\\
&+& 2 |h_{\alpha 2}|^2 |S_{22}(p)|^2{\rm Re}\left(h_{\beta 1}^{\star} h_{\beta 2} S_{21}^{\star}(p') S_{22}(p')\right)\nonumber\\
&+& 2 |h_{\beta 1}|^2 |S_{11}(p')|^2{\rm Re}\left(h_{\alpha 1}^{\star} h_{\alpha 2} S_{11}^{\star}(p) S_{12}(p)\right)\nonumber\\
&+& 2 |h_{\beta 2}|^2 |S_{22}(p')|^2{\rm Re}\left(h_{\alpha 1}^{\star} h_{\alpha 2} S_{21}^{\star}(p) S_{22}(p)\right)\nonumber\\
&+& 2 |h_{\alpha 1}|^2{\rm Re}\left(h_{\beta 1}^{\star} h_{\beta 2} S_{11}^{\star}(p) S_{21}(p)S_{11}^{\star}(p') S_{22}(p')\right)\nonumber\\
&+& 2 |h_{\alpha 2}|^2{\rm Re}\left(h_{\beta 1}^{\star} h_{\beta 2} S_{12}^{\star}(p) S_{22}(p)S_{11}^{\star}(p') S_{22}(p')\right)\nonumber\\
&+& 2 |h_{\beta 1}|^2{\rm Re}\left(h_{\alpha 1}^{\star} h_{\alpha 2} S_{11}^{\star}(p) S_{22}(p)S_{11}^{\star}(p') S_{21}(p')\right)\nonumber\\
&+& 2 |h_{\beta 2}|^2{\rm Re}\left(h_{\alpha 1}^{\star} h_{\alpha 2} S_{11}^{\star}(p) S_{22}(p)S_{12}^{\star}(p') S_{22}(p')\right).\hspace{0.5cm}\label{exp}
\end{eqnarray}
The first two terms are the ones that are naively expected if the RH neutrino propagation
is incoherent. In the on-shell limit, within a narrow width approximation, they yield, respectively,
\begin{eqnarray}
|S_{ii}(p)|^2&\propto&{1\over p^2-s_i }{1\over p^2-s_i^{\star}} \nonumber\\
&\longrightarrow&{\pi\over \Gamma_i M_i}\, \delta(p^2-M_i^2)\, . \label{incoh}
\end{eqnarray}
They correspond to two incoherent RH neutrinos produced in $Z'$ decay which subsequently decay into a lepton-Higgs pair.
The third term in Eq.~(\ref{exp}) is new. It is an interference term that does not rely on the mixing,
or alternatively on the
off-diagonal elements of the matrix of propagators. Let us analyze this term in detail:
\begin{eqnarray}
S_{11}^{\star}(p)S_{22}(p)&\propto& {1\over p^2-s_1^{\star}}{1\over p^2-s_2} \nonumber\\
&& \hspace{-1.5cm}\longrightarrow
{2\pi i\over s_2-s_1^{\star}}\delta(p^2-\overline{M}^2)\nonumber\\
&&\hspace{-1.5cm}= {2\pi i(M_2^2-M_1^2+i(\Gamma_1 M_1+\Gamma_2M_2))\over (M_2^2-M_1^2)^2+
(\Gamma_1 M_1+\Gamma_2 M_2)^2}\delta(p^2-\overline{M}^2)\, , \hspace{0.5cm}\label{s11s22}
\end{eqnarray}
where $\overline{M}^2\equiv (M_1^2+M_2^2)/2$.
A similar term arises from $S_{11}^{\star}(p')S_{22}(p')$. Dropping the delta functions and the $\pi$ factors,
we are left with the evaluation of
\begin{eqnarray}
&&\hspace{-1cm}{\rm Re}\left\{ {-\left[M_2^2-M_1^2+i(\Gamma_1 M_1+\Gamma_2 M_2)\right]^2\over
\left[(M_2^2-M_1^2)^2+(\Gamma_1 M_1+\Gamma_2 M_2)^2\right]^2} h_{\alpha 1}^{\star}h_{\beta 1}^{\star} h_{\alpha 2}h_{\beta 2}\right\}\nonumber\\
&&\hspace{-1cm}= {-(M_2^2-M_1^2)^2+(\Gamma_1 M_1+\Gamma_2 M_2)^2\over
\left[(M_2^2-M_1^2)^2+(\Gamma_1 M_1+\Gamma_2 M_2)^2\right]^2}{\rm Re}(h_{\alpha 1}^{\star}h_{\beta 1}^{\star} h_{\alpha 2}h_{\beta 2})+\nonumber\\
&&\hspace{-1cm} {2(M_2^2-M_1^2)(\Gamma_1 M_1+\Gamma_2 M_2)\over
\left[(M_2^2-M_1^2)^2+(\Gamma_1 M_1+\Gamma_2 M_2)^2\right]^2}{\rm Im}(h_{\alpha 1}^{\star}h_{\beta 1}^{\star} h_{\alpha 2}h_{\beta 2}).\label{third}
\end{eqnarray}

For the rate into two antileptons, $|\mathcal{M}(\bar{\ell}_{\alpha}\bar{\ell}_{\beta})|^2$,
only the replacement $h_{\alpha i}\to h_{\alpha i}^{\star}$ needs to be made in Eq.~(\ref{third}). The difference
in the rates into two leptons and two antileptons is then given by
\begin{equation}\label{supp}
|\mathcal{M}(\ell_{\alpha}\ell_{\beta})|^2-|\mathcal{M}(\bar{\ell}_{\alpha}\bar{\ell}_{\beta})|^2 \propto
\epsilon_+^2 \,{\Gamma_1 M_1+\Gamma_2 M_2\over M_2^2 -M_1^2}\,,
\end{equation}
where we introduced the parameter
\begin{equation}
\epsilon_{\pm}  \equiv{(M_2^2-M_1^2)M_2 \Gamma_2\over
(M_2^2-M_1^2)^2+(\Gamma_1 M_1\pm \Gamma_2 M_2)^2}.
\end{equation}
Note that in the following we shall assume a hierarchy between $M_1\Gamma_1$ and $M_2\Gamma_2$ such that
$\epsilon_+\simeq \epsilon_-=\epsilon$.
In the limit $M_2-M_1 \gg \Gamma /2$, the overall suppression of the third term in Eq.~(\ref{exp})
is therefore $\epsilon^3$. As for the sum of the rates
$|\mathcal{M}(\ell_{\alpha}\ell_{\beta})|^2+|\mathcal{M}(\bar{\ell}_{\alpha}\bar{\ell}_{\beta})|^2$ it can be
readily seen that it is suppressed as $\epsilon^2$.

For future use, we note that the
$C\!P$ asymmetry in resonant leptogenesis, summed over flavor, is given by~\cite{Anisimov:2005hr}
\begin{equation}\label{CPres}
\varepsilon_1={{\rm Im}(h^{\dagger}h)_{12}^2\over (h^{\dagger}h)_{11}(h^{\dagger}h)_{22}} \,\epsilon_-\,.
\end{equation}

Terms 4 to 7 in Eq.~(\ref{exp}) involve mixing. The prefactors $|S_{ii}(p)|^2$ yield
delta functions as in Eq.~(\ref{incoh}). The more interesting part is the real part. We have
\begin{eqnarray}
S_{11}^{\star}(p')S_{12}(p')&=&{1\over p'^2-s_1^{\star}}\times\nonumber\\
&&\hspace{-1cm}{M_1 M_2 \Sigma_{12}^R +p'^2 \Sigma_{21}^R
+M_1 \Sigma_{12}^{M\star} +M_2 \Sigma_{12}^M\over (p'^2-s_1)(p'^2-s_2)}\nonumber\\
&=& {\pi \delta(p'^2-M_1^2)\over \Gamma_1 M_1} \, X_{12}\,,\label{S11S12}
\end{eqnarray}
where $\Sigma^R$ and $\Sigma^M$ are defined in \cite{Anisimov:2005hr} [Eqs.~(9)--(12)], and
with
\begin{equation}\label{x12}
X_{12}\equiv {M_1 M_2 \Sigma_{12}^R +M_1^2 \Sigma_{21}^R
+M_1 \Sigma_{12}^{M\star} +M_2 \Sigma_{12}^M\over (s_1-s_2)}.
\end{equation}
To evaluate ${\rm Re}\left(h_{\beta 1}^{\star} h_{\beta 2}
X_{12}\right)$, we multiply first the numerator and the denominator
by $(s_1-s_2)^{\star}$, noting that
\begin{equation}
|s_1-s_2|^2= (M_1^2-M_2^2)^2+(\Gamma_1 M_1-\Gamma_2 M_2)^2.
\end{equation}
Then we turn to the rate into antileptons and find it to be proportional to
${\rm Re}\left(h_{\beta 1} h_{\beta 2}^{\star} X_{21}\right)$, where $X_{21}$ is equal
to $X_{12}$ modulo the transformation of the Yukawas $h\to h^{\star}$. The difference
$|\mathcal{M}(\ell_{\alpha}\ell_{\beta})|^2-|\mathcal{M}(\bar{\ell}_{\alpha}\bar{\ell}_{\beta})|^2$,
summed over $\beta$ for convenience, is then found to be precisely proportional to
$\varepsilon_1$ as defined in Eq.~(\ref{CPres}).

The fifth term in Eq.~(\ref{exp}) will give another $\varepsilon_1$ contribution
(from the other leg), whereas
the terms 6 and 7 in Eq.~(\ref{exp}) are the $\varepsilon_2$ contributions.

Now let us turn to terms 8 to 11 in Eq.~(\ref{exp}).
We will only show explicitly how to work out term 8, but
the other terms follow in a similar fashion. From Eqs.~(\ref{s11s22}) and~(\ref{S11S12})
we have that
\begin{eqnarray}
&&{\rm Re}\left[h_{\beta 1}^{\star} h_{\beta 2} S^{\star}_{11}(p) S_{21}(p)S^{\star}_{11}(p') S_{22}(p')\right]=\nonumber\\
&&{\rm Re}\left[h_{\beta 1}^{\star} h_{\beta 2} S^{\star}_{11}(p) S_{21}(p)\right]{\rm Re}\left[S^{\star}_{11}(p') S_{22}(p')\right]
\nonumber\\
&&-{\rm Im}\left[h_{\beta 1}^{\star} h_{\beta 2} S^{\star}_{11}(p) S_{21}(p)\right]{\rm Im}\left[S^{\star}_{11}(p') S_{22}(p')\right]
\nonumber\\
&&\propto{2\pi^2\over M_1 \Gamma_1} \delta(p^2-M_1^2)\delta(p'^2-M_2^2)\times \nonumber\\
&&\left[{\rm Re}\left(h_{\beta 1}^{\star} h_{\beta 2} X_{12}\right)
{M_1\Gamma_1+M_2\Gamma_2\over |s_2-s_1^{\star}|^2}\right.\nonumber\\
&&\left.\hspace{1cm}- {\rm Im}\left(h_{\beta 1}^{\star} h_{\beta
2}X_{12}\right) {M_2^2-M_1^2\over |s_2-s_1^{\star}|^2}\right]\label{mixed}
\end{eqnarray}
When estimating the difference of rates
into leptons and antileptons, we have already seen that ${\rm Re}\left(h_{\beta 1}^{\star}
h_{\beta 2}X_{12}\right)$ gives rise to a term proportional to
the $C\!P$ asymmetry $\varepsilon_1$. But here it is multiplied by
another small term in the limit $M_2-M_1 \gg \Gamma /2$, so that the
overall suppression is equivalent to Eq.~(\ref{supp}), namely an
$\epsilon^3$ suppression.

As for the second term in the square bracket, there is an obvious suppression by
$\epsilon$ to start with. The difference of
rates into leptons and antileptons yields another suppression. It is given by
\begin{eqnarray}
{\rm Im}\left(h_{\beta 1}^{\star} h_{\beta 2}X_{12}\right)&-&{\rm
Im}\left(h_{\beta 2}^{\star} h_{\beta 1}X_{21}\right)=\nonumber\\
&&\hspace{-1cm}{\rm
Im}\left(h_{\beta 1}^{\star} h_{\beta 2}\right)\left({\rm Re}X_{12}+
{\rm Re} X_{21}\right)\nonumber\\
&&\hspace{-1.2cm}+{\rm Re}\left(h_{\beta 1}^{\star} h_{\beta 2}\right)\left({\rm
Im}X_{12}-{\rm Im}X_{21}\right).
\end{eqnarray}
The first term on the right-hand side is given to leading order by
\begin{eqnarray}
{\rm Re}X_{12}+{\rm Re}X_{21}&\simeq&\left({\rm Im}\Sigma_{12}+{\rm
Im}\Sigma_{21}\right){M_2 \Gamma_2-M_1 \Gamma_1\over
|s_2-s_1|^2}\nonumber
\\
&&\hspace{-2cm}={M_1(M_1+M_2)\over 8\pi}{\rm Re}(h^{\dagger}h)_{21}{M_2
\Gamma_2-M_1 \Gamma_1\over |s_2-s_1|^2},
\end{eqnarray}
which implies an additional suppression by $\epsilon^2$, such that
the overall suppression is $\epsilon^3$. As before, the sum of rates into
leptons and antileptons is only suppressed by a factor $\epsilon^2$.

The second term
yields
\begin{eqnarray}
{\rm Im}X_{12}-{\rm Im}X_{21}&\simeq&\left({\rm Re}\Sigma_{12}-{\rm
Re}\Sigma_{21}\right){\Gamma_2M_2-\Gamma_1 M_1\over
|s_2-s_1|^2}\nonumber
\\
&&\hspace{-2cm}={M_1(M_2-M_1)\over 8\pi}{\rm
Im}(h^{\dagger}h)_{12}{\Gamma_2M_2-\Gamma_1 M_1\over
|s_2-s_1|^2},
\end{eqnarray}
which is even more suppressed than the previous term.

We have therefore
shown that the $C\!P$-violating interference effects are at least suppressed
by three powers of $\epsilon$, while the $C\!P$ conserving ones are suppressed
by two powers only.

\subsection{Oscillation derivation}

We employ here the formalism which was successfully used to describe $K^0$--$\overline{K}^0$ and $B^0$--$\overline{B}^0$
oscillations. The first part of our discussion will follow closely~\cite{Covi:1996fm}, where leptogenesis from mixed particle decays
was considered.

In the non-relativistic limit the squared Hamiltonian can be decomposed as
\begin{equation}
\hat{H}_{ij}=M_{ij}-{\rm i}\,\Gamma_{ij}/2
\end{equation}
where the renormalized mass matrix $M$ includes the dispersive parts of the self-energy diagram while the matrix $\Gamma$
arises from the absorptive part alone. The squared Hamiltonian can be diagonalized by a non-unitary matrix $V$,
\begin{equation}
(V\,\hat{H}\, V^{-1})_{ij}=\sqrt{s_i}\delta_{ij}\,,
\end{equation}
and the squared eigenvalues of the Hamiltonian coincide with the poles of the propagator defined in Eq.~(\ref{poles}).

An important ingredient in the present formalism is the proper identification of the initial state. In our case,
the $Z'$ is produced in the $s$ channel, and then decays into a pair of RH neutrinos. The only gauge invariant
combination is $|N'_1\overline{N}'_1+N'_2\overline{N}'_2\rangle$, where $N'_i$ are propagation
eigenstates. The decay rate into
dileptons will then be proportional to
\begin{equation}
|\mathcal{M}(\ell_{\alpha}\ell_{\beta})|^2=|\langle \ell_{\alpha}\Phi \ell_{\beta}\Phi| H_{\rm int}^2|
N'_1\overline{N}'_1+N'_2\overline{N}'_2\rangle|^2\,.
\end{equation}
Expanding this expression we obtain
\begin{eqnarray}
|\mathcal{M}(\ell_{\alpha}\ell_{\beta})|^2&=&|\langle \ell_{\alpha} \Phi|H_{\rm int}|N'_1\rangle|^2
|\langle \ell_{\beta} \Phi|H_{\rm int}|N'_1\rangle|^2\nonumber\\
&+&|\langle \ell_{\alpha} \Phi|H_{\rm int}|N'_2\rangle|^2
|\langle \ell_{\beta} \Phi|H_{\rm int}|N'_2\rangle|^2\nonumber\\
&+&2\,{\rm Re}\left(\langle \ell_{\alpha} \Phi|H_{\rm int}|N'_2\rangle\langle \ell_{\beta} \Phi|H_{\rm int}|N'_2\rangle\right.\nonumber\\
&&\left.\langle N'_1|H_{\rm int}^{\dagger}|\ell_{\alpha}\Phi\rangle\langle N'_1|H_{\rm int}^{\dagger}|\ell_{\beta}\Phi\rangle\right)\,.
\label{oscilexp}
\end{eqnarray}
Evolving the propagation eigenstates in the usual way, we have that
\begin{eqnarray}
T_{i\beta}(t)&\equiv& \langle \ell_{\beta} \Phi|H_{\rm int}|N'_i(t)\rangle\nonumber\\
&=&(h_{\beta 1}V_{1i}^{-1}+h_{\beta 2}V_{2i}^{-1})\,e^{-{\rm i}\,\sqrt{s_i} t} .
\end{eqnarray}
Let us now compute the first term in Eq.~(\ref{oscilexp}) allowing for oscillation,
namely $|T_{1\alpha}(t)|^2$ to first
order in the off-diagonal elements of the mixing matrix $V$. Note that
$V_{11}=V_{22}=1+\mathcal{O}(h^2/(16\pi^2))$. We obtain
\begin{eqnarray}
|T_{1\beta}(t)|^2&=&|h_{\beta 1}|^2 e^{{\rm i}(\sqrt{s_1}^{\star}-\sqrt{s_1})t}\nonumber\\
&+&2\,{\rm Re}\left(h_{\beta 1}^{\star}h_{\beta 2}
V_{21}^{-1}e^{{\rm i}(\sqrt{s_1}^{\star}-\sqrt{s_1})t}\right)\,.
\end{eqnarray}
Carrying out the time integration from 0 to infinity, we have
\begin{eqnarray}
\int_0^{\infty}dt\,|T_{1\beta}(t)|^2&=&|h_{\beta 1}|^2 {1\over\Gamma_1} \nonumber\\
&+&2{1\over\Gamma_1}{\rm Re}\left(h_{\beta 1}^{\star}h_{\beta 2}
V_{21}^{-1}\right)\label{T1beta}
\end{eqnarray}
and $V_{21}^{-1}=-V_{21}$ coincides with $X_{12}$ defined in Eq.~(\ref{x12}).

We obtain the first term in Eq.~(\ref{oscilexp}) by
multiplying with the same expression as derived in Eq.~(\ref{T1beta})
except for the replacement $\beta\to \alpha$. To first order in $V_{21}$,
we obtain five terms, which correspond to the first, fourth, sixth term in Eq.~(\ref{exp}).
Note that the factor $g_1'^2\pi/M$ difference between this formalism
and the previous one can be trivially explained when computing explicitly
the total cross-section
$q\bar{q}\to \ell_{\alpha}\ell_{\beta}\Phi\Phi$.

It is easy to obtain the corresponding expressions for $|T_{2\beta}(t)|^2$, and the second
term in Eq.~(\ref{oscilexp}) readily yields terms 2, 5, 7 in Eq.~(\ref{exp}).

For the third term in Eq.~(\ref{oscilexp}) we have that
\begin{eqnarray}
T_{1\beta}^{\star}(t)T_{2\beta}(t)&=&h_{\beta 1}^{\star}h_{\beta 2} e^{{\rm i}(\sqrt{s_1}^{\star}-\sqrt{s_2})t}\nonumber\\
&+& |h_{\beta 1}|^2\, V_{12}^{-1}e^{{\rm i}(\sqrt{s_1}^{\star}-\sqrt{s_1})t}\nonumber\\
&+& |h_{\beta 2}|^2\, (V_{21}^{-1})^{\star} e^{{\rm i}(\sqrt{s_2}^{\star}-\sqrt{s_2})t}
\end{eqnarray}
to first order in $V_{21}$, and the time integration yields
\begin{eqnarray}
\int_0^{\infty}dt\, T_{1\beta}^{\star}(t)T_{2\beta}(t)&=&h_{\beta 1}^{\star}h_{\beta 2}{{\rm i}\,2M_1\over s_2-s_1^{\star}}\nonumber\\
&+& {1\over \Gamma_1}|h_{\beta 1}|^2\, V_{12}^{-1}\nonumber\\
&+& {1\over \Gamma_2}|h_{\beta 2}|^2\, (V_{21}^{-1})^{\star}\,.
\end{eqnarray}
A similar result can be obtained for $T_{1\alpha}^{\star}(t)T_{2\alpha}(t)$. Multiplying the two expressions, we
obtain the third term in Eq.~(\ref{exp}), as well as terms 8 to 11.
This completes the proof of the equivalence between the field theory
formalism and the oscillation one.

\section{A Framework for Natural Resonant Leptogenesis}

In this section, we outline a framework which naturally realizes resonant
leptogenesis at the TeV scale. In particular, the following features that
are necessary for the scenario we have proposed to be viable will be shown to
emerge naturally in this scheme.
 \begin{itemize}
 \item
A simple understanding of why the RH neutrino masses naturally lie
close to the weak scale.
 \item
A straightforward explanation for the smallness of neutrino masses, and
for the quasi-degeneracy of the RH neutrinos.
 \item
A natural understanding of why the asymmetry in RH neutrino
decays tends to be of order one.
 \end{itemize}

We begin by extending the electroweak gauge group of the SM from
$SU(2)_L\times U(1)_Y$ to $SU(2)_L\times U(1)_Y\times U(1)_{B-L}$. The
assignment of the SM fermion charges under this gauge group is obvious,
being dictated by their $B-L$ charge. We stress that this choice of
charges is motivated primarily by simplicity, and that a different choice
would not affect our conclusions. Anomalies are cancelled by three right
handed neutrinos $N_i$, which are each SM singlets. In our framework, this
extended gauge symmetry is broken to that of the SM by a scalar $\Delta$
with $B-L$ charge +2 that breaks $U(1)_{B-L}$ completely. The Yukawa
coupling of $\Delta$ to the RH neutrinos gives them masses at the $B-L$
breaking scale. The interactions of the RH neutrinos take the form below
 \begin{equation}
{\cal L}~=~\sum_i \sum_j \left[ f^{ij} \Delta N_i N_j +
h^{ij} L_i H N_j \right]
 \end{equation}
For weak scale resonant leptogenesis we require $h \sim 10^{-6}$, $\langle
\Delta \rangle \sim$ 1 TeV and $f^{ij} = f \delta^{ij}$ up to corrections
of order $h^2/ 8 \pi$. The challenge before us is to explain these
features.

A simple understanding of the smallness of the Dirac neutrino couplings
$h^{ij}$ can be obtained through the extra dimensional `shining
mechanism'~\cite{AHTW}. Consider a five dimensional theory, with the extra
dimension $y$ compactified on $S^1/Z_2$. The radius of the extra dimension is
denoted by $R$, and there are branes at the orbifold fixed points $y = 0$ and
$ y = \pi R$. The extra dimension is assumed to be extremely small, so that
the compactification scale $R^{-1}$ is much larger than the TeV scale, and is
of order the grand unification scale or higher, $R^{-1} \geq 10^{16}$ GeV. All
the SM fields are localized on the brane at $y =\pi R$, while the RH neutrinos
$N_i$ and the $B - L$ gauge boson occupy the bulk of the space. The field
$\Delta$ which breaks the $B - L$ symmetry is localized to the brane at $ y =
0$.

We now outline how the various interactions arise in this scheme. In order
to naturally resolve the hierarchy problem of the SM, we work in a
supersymmetric framework. The SM matter and Higgs fields are promoted to
four dimensional chiral superfields while the SM gauge fields become
components of four dimensional vector superfields. The field $\Delta$ also
becomes part of a four dimensional chiral supefield, and must be
complemented by a separate chiral superfield $\overline{\Delta}$ to ensure
anomaly cancellation. On the other hand, the RH neutrinos must
be incorporated into a hypermultiplet in five dimensions, while the $B -
L$ gauge boson is now part of a five dimensional gauge multiplet.

To specify the boundary conditions to be satisfied by bulk fields we need
to know their transformation properties under reflections about $y=0$,
which we denote by $Z$. In addition, we also need to specify either their
transformation properties under translations by $2 \pi R$, which we denote
by $T$, or their transformation properties under reflections about $\pi
R$, which we denote by $Z'$. $T$ and $Z'$ are related by $Z' = T \; Z$. We
choose to describe the boundary conditions satisfied by the various fields
in terms of $Z$ and $Z'$.

A supersymmetric vector multiplet $\hat{V}$ in five dimensions consists of
a five dimensional gauge field $A_M$, an adjoint scalar $\sigma$, and
fermionic fields $\lambda$ and $\lambda'$. From the four dimensional
viewpoint the five dimensional theory has $\mathcal{N}=2$ supersymmetry.
Under the action of $Z$ and $Z'$ this $\mathcal{N}=2$ supersymmetry is
broken to $\mathcal{N}=1$ supersymmetry. The five dimensional multiplet
can be broken up into four dimensional $\mathcal{N}=1$ supermultiplets as
$\hat{V} = \left(V, \Sigma \right)$ where the vector multiplet $V$
consists of $\left(A_{\mu}, \lambda \right)$ and the chiral multiplet
$\Sigma$ consists of $\left( \sigma + i A_5, \lambda' \right)$. $V$ and
$\Sigma$ must necessarily have different transformation properties under
$Z$. In order to obtain a light zero mode for the $B - L$ gauge boson we
assign $V$ in the corresponding five dimensional gauge multiplet a parity
of +1 under both $Z$ and $Z'$, while $\Sigma$ is assigned a parity of -1.

A hypermultiplet $\hat{N}$ in five dimensions consists of bosonic fields
$\tilde{N}$ and $\tilde{N}^c$ and fermionic fields $N$ and $N^c$. The
hypermultiplet can be decomposed into four dimensional $\mathcal{N}=1$
superfields. Then $\hat{N}$ breaks up into $(n, n^c)$ where $n =
(\tilde{N}, N)$ and $n^c =(\tilde{N}^c, N^c)$. Since $n$ and $n^c$ have
different transformation properties under $Z$, the four dimensional
$\mathcal{N}=2$ supersymmetry of the system is broken to $\mathcal{N}=1$.
Since we require the RH neutrinos to have zero modes we assign
$n$ in each of the corresponding five dimensional hypermultiplets a parity
of +1 under both $Z$ and $Z'$, while $n^c$ is assigned a parity of -1.

The bulk action for the RH neutrinos, in a formalism which
keeps four dimensional $\mathcal{N} = 1$ supersymmetry
manifest~\cite{AGW}, takes the
form
below.
 \begin{eqnarray}
S =
\int d^4 x dy \left[ \int d^4 \theta
\left( {n_i}^{\dagger} e^{V} n_i + {n_i^c}^{\dagger} e^{-V} n^c \right)
\right.
\nonumber
\\\left.
+
\int d^2 \theta n_i^c \left( m_i + \partial_y \right) n_i \right]
 \end{eqnarray}
The mass term $m$ is odd under the $Z_2$ symmetry $y \rightarrow -y$,
and therefore does not contribute to the mass of the zero modes. Its
effect is to give the zero modes of $n$ a profile $\sim e^{-my}$, which
is exponentially localized towards $y = 0$ for $m > (\pi R)^{-1}$.

The interactions of RH neutrinos are now localized on the branes
and take the form below
 \begin{equation}
{\cal L}~=~
\int d^2 \theta \delta( y) \sum_{i,j} \hat{f}^{ij} \Delta n_i n_j
+ \delta(y - \pi R) \sum_{i,j} \hat{h}^{ij} L_i H n_j
 \end{equation}
We now impose an $SO(3)_H$ horizontal symmetry which rotates the $n_i$
into each other, and also the $n_i^c$ into each other. This symmetry is
exact in the bulk, and on the brane at $y = 0$, but is assumed to be
broken on the brane at $y = \pi R$. Then the bulk mass term $m_i = m$ and
the coupling $\hat{f}^{ij} = \hat{f} \delta^{ij}$. However, the coupling
$\hat{h}^{ij}$ retains non-trivial flavor structure. Then, after
normalizing appropriately the interactions of the zero mode RH neutrino
superfields in the four dimensional effective theory below the
compactification scale take the form
 \begin{equation}
\int d^2 \theta f \Delta n_i n_i + \sum_{i,j} {h}^{ij} L_i H n_j
 \end{equation}
Provided the compactification scale $R^{-1}$ is not far from the cutoff
$\Lambda$ of the higher dimensional theory, $\Lambda R \lesssim 10$, then
the coupling constant $f$ can be order one. However, the exponential
profile of the zero mode RH neutrino fields means that the Dirac Yukawa
coupling $h$ is exponentially suppressed, $h \sim {\rm exp} \left( - m \pi
R \right)$. For $\Lambda R$ of order a few, we then have a natural
understanding of why the Dirac Yukawa couplings in the neutrino sector
are small.

As we now explain this framework also leads to a natural understanding of
why $\langle \Delta \rangle \sim$ TeV, along the lines of radiative
electroweak breaking in the Minimal Supersymmetric Standard Model (MSSM).
Let the scalar superpartners of the $n_i$ acquire a soft supersymmetry
breaking mass $\widetilde{m}_n$ of order a TeV
through any of several mediation mechanisms, such as anomaly
mediation~\cite{RS0} or gaugino mediation~\cite{KKS}. Then, if the Yukawa
couplings $f_i$ are of order one, there is a large logarithmically
enhanced negative contribution to the soft mass of $\Delta$,
 \begin{equation}
\delta \widetilde{m}^2_{\Delta} \approx - \frac{3 f_i^2}{8 \pi^2}
\widetilde{m}^2_n {\rm log}\left( \frac{\Lambda^2}{\widetilde{m}^2_n}
\right)
 \end{equation}
The logarithmic enhancement means that this radiative contribution can
naturally dominate over a comparable positive tree-level contribution to
the soft mass of $\Delta$, leading to dynamical breaking of the $B - L$
symmetry. The quartic that stabilizes $\langle \Delta \rangle$ is provided
by the D-term of $U(1)_{B-L}$, leading to
 \begin{equation}
\langle \Delta \rangle \approx
\sqrt{\frac{|\widetilde{m}^2_{\Delta}|}{g_{B - L}^2}}
 \end{equation}
Then the right handed scale and the $B - L$ gauge boson mass are both
naturally of order a TeV, exactly in the right range to generate a signal
at the LHC.

Finally we turn to the asymmetry parameter $\epsilon$. To estimate
$\epsilon$ we first consider the mass splitting of the RH
neutrinos. Logarithmically enhanced radiative effects arising from the
Dirac Yukawa coupling $h$ lead to a small splitting in the coupling
constant $f$ for the different generations
\begin{equation}
\delta f \sim \frac{h^2}{16 \pi^2} {\rm log}
\left( \frac{\Lambda^2}{\widetilde{m}^2_n} \right)
\end{equation}
This in turn breaks the degeneracy of the RH neutrinos,
$\delta m_n \sim \delta f \langle \Delta \rangle$. This must be
compared to the decay width of the RH neutrinos
\begin{equation}
\Gamma \sim \frac{h^2}{8 \pi} m_n
\end{equation}
From the ratio $\Gamma/\delta m_n$ we see that the natural values
of $\epsilon$ are indeed in the neighbourhood of 0.1, as required
for successful weak scale leptogenesis.

In summary, we see that the framework outlined here can naturally
explain the features necessary for a successful theory of resonant
leptogenesis, without any need for the fine tuning of parameters.

\end{document}